\documentclass[12pt]{article}

\usepackage{graphicx}
\usepackage{epsfig}
\usepackage[dvips]{color}

\newcommand{\be}{\begin{eqnarray}}
\newcommand{\ee}{\end{eqnarray}}

\begin{document}

\hbox{} \nopagebreak \vspace{-3cm}
\addtolength{\baselineskip}{.8mm} \baselineskip=24pt


\begin{flushright}
{\sc BI-TP 2003/12}\\
\end{flushright}
\vspace{-1.0cm}
\begin{flushright}
{\sc CERN-TH 2003/116}\\
\end{flushright}

\vspace{40pt}

\begin{center}
{\large {\sc {\bf  Saturation and parton level Cronin effect:
enhancement {\it vs} suppression
of gluon production in p-A and A-A collisions}}}\\
\baselineskip=12pt \vspace{34pt}
Rudolf Baier $^{1,4}$, Alex Kovner$^{2,4}$ and Urs Achim Wiedemann $^{3,4}$
\vspace{24pt}

$^1$Fakult\"at f\"ur Physik, Universit\"at Bielefeld,
D-33615 Bielefeld, Germany\\[0pt]
$^2$ Department of Mathematics and Statistics, University of
Plymouth, Drake Circus,
Plymouth PL4 8AA, UK\\[0pt]
$^3$ Theory Division, CERN, CH-1211 Geneva 23, Switzerland\\[0pt]
$^4$ Institute for Nuclear Theory, University of Washington, Box
351550, Seattle, WA 98195, USA
 \vspace{60pt}

\end{center}

\vspace{40pt}

\begin{abstract}
We note that the phenomenon of perturbative saturation leads to
transverse momentum broadening in the spectrum of partons produced
in hadronic collisions. This broadening has a simple
interpretation as parton level Cronin effect
for systems in which saturation is generated by the "tree level"
Glauber-Mueller mechanism. For systems where the broadening results
form the nonlinear QCD evolution to high energy, the presence or absence of 
Cronin effect depends crucially on the quantitative behavior of the
gluon distribution functions at transverse momenta $k_t$ outside the 
so called scaling window. We discuss the
relation of this phenomenon to the recent analysis by
Kharzeev-Levin-McLerran of the momentum and centrality dependence 
of particle production in nucleus-nucleus collisions at
RHIC.
\end{abstract}

\vfill

\newpage

\section{Introduction}

The phenomenon of perturbative saturation has been the focus of
intensive study in recent years. Since the appearance of the first
RHIC data, 
saturation based ideas have motivated several attempts
at understanding bulk properties of ultra-relativistic heavy ion
collisions such as the multiplicity, rapidity distribution and
centrality dependence of particle production~\cite{kln,klm}. 
In particular the recent work~\cite{klm}
suggests that the saturating properties of the nuclear gluon
distribution may be responsible for the $N_{\rm part}$
scaling of charged particle multiplicity  at $2<p_t<8$ GeV with
centrality in RHIC data. It is undoubtedly true that saturation
effects suppress the number of gluons below the saturation
momentum $Q_s$ in the nuclear wave function. It is then very
plausible that the number of produced particles at $p_t<Q_s$ is
also suppressed relative to simple perturbative prediction. The
value of $Q_s^2$ is estimated to be of order $2$ GeV$^2$ for most
central collisions at RHIC~\cite{kln}. The suggestion of~\cite{klm}
goes beyond this simple statement and implies that the suppression
persists also at higher momenta, namely in the so called scaling
window $Q_s<p_t<Q_s^2/Q_0$.

On the other hand one expects a competing effect of transverse
momentum broadening due to multiple rescatterings in the final
state, the so called Cronin effect. This should enhance the number
of particles produced in the intermediate momentum range and thus
works against the saturation argument. The purpose of
this note is to show that there is no outright contradiction between the
appearance of the Cronin effect and the expectations based on the
saturation scenario and that, in fact, the Cronin enhancement is
inherent in some realizations of 
saturation physics. We show here that saturation models based on multiple
rescattering lead to the relation
\begin{equation}
  {dN^{\rm sat}\over d^2p_t} > {dN^{\rm pert}\over d^2p_t}
  \label{eq1}
\end{equation}
for all $p_t>Q_s$. 
This relation leads to Cronin enhancement when comparing gluon production
in central and peripheral collisions.

For saturation models based not on multiple rescattering but rather on coherent
suppression at low $x$ the situation is more complicated. 
Here it is also true that the number of gluons produced in the intermediate
momentum range is greater
than the prediction of the leading order perturbation theory. This overall
enhancement is the result of the increase of the number of gluons in the
nuclear wave function due to low $x$ evolution.
The transverse 
momentum broadening is also present. This is manifested by the
anomalous dimension generated by the low $x$ evolution, so that the decrease of
the evolved distribution with momentum is very slow.
However, whether this broadening
results in the Cronin enhancement is determined by the 
behavior of saturated gluon distribution at high transverse momenta. The 
momenta which are important are those above the
scaling window inside which the value of the anomalous dimension is 
analytically understood. 
If the distribution above the scaling window approaches quickly the leading 
order perturbative one, the Cronin effect is indeed generated. However if the
distribution continues to vary much slower than $1/p_t^2$, the (properly
normalized) multiplicity
of produced gluons is always smaller for central collisions than for the 
peripheral ones. 

In this light, the results of ~\cite{klm} should be
understood in the following way. In a particular saturation ansatz
considered in ~\cite{klm} for $p_t$ in the scaling window,
${dN^{\rm sat}\over d^2p_t}$ as a function of centrality is indeed
proportional to the number of participants $N_{\rm part}$. However
its value is always {\it greater} than that for the leading order perturbative
prediction for the same momentum, even though the latter is
proportional to the number of collisions $N_{\rm coll}\propto
N^{4/3}_{\rm part}$. This holds as long as the transverse momentum
in question is greater than the saturation momentum for the most
central collisions considered. Thus according to the saturation
scenario, the number of produced particles in the intermediate
transverse momentum range is enhanced and not suppressed relative
to the leading order perturbative one.
This enhancement in the overall production may or may not result in the Cronin
enhancement when comparing the production rates for central versus 
peripheral collisions, as the evolution is equally effective for
all impact parameters.

Before we proceed further, we wish to make clear the following
points. First, to calculate the gluon production we use the $k_t$
factorized formalism as in~\cite{kln,klm}. Although it has
not been proved to hold for this process and is likely not to be
strictly valid, one may hope as in ~\cite{kln} that it gives a
qualitatively reasonable description of gluon production. Second,
in this note we only address the multiplicity of produced gluons.
For comparison with experimental data, this quantity has to be
convoluted with gluon fragmentation functions to convert it to 
the number of produced hadrons. This introduces additional 
uncertainties related to our limited knowledge of the dynamics 
of the system between the time of production and the time of
hadronization~\cite{bmss}, and the possible medium-dependence of
parton fragmentation ~\cite{sw}. Thus our results have to be
paralleled to those of ~\cite{klm} prior to convolution with 
fragmentation functions.

The simple $k_t$-factorized formula for the gluon yield at central
rapidity in a collision of identical nuclei ~\cite{glr} used in
~\cite{klm} is
\begin{equation}
   E{d\sigma\over d^3p\, d^2b}
   ={4\pi^2\alpha_s S_{AA}(b)N_c\over N_c^2-1}{1\over p_t^2}
    \int d^2k_t\, \phi_A(y,k_t)\, \phi_A(y,p_t-k_t)\, .
 \label{eq2}
\end{equation}
Here $S_{AA}(b)$ is the overlap area in the transverse plane
between the nuclei at fixed impact parameter $b$, $y$ is the
rapidity difference between the central rapidity and the
fragmentation region and $\phi_A(y,k_t)$ is the intrinsic
momentum dependent nuclear gluon distribution function, related to
the standard gluon distribution by
\begin{equation}
  \phi_A(y,k_t)={d(xG_A(x,k^2_t))\over d^2k_t\, d^2b}\, .
  \label{eq3}
\end{equation}
Eq.~(\ref{eq2}) itself is an approximation even within the
$k_t$-factorization scheme for two reasons. First, the gluon
distribution is considered to be effectively impact parameter
independent and taken at some representative impact parameter
inside the overlap region. Second the rapidity of both
distribution functions is taken to be the same rather than
integrating over the relative rapidity of the two distributions.
This amounts to the assumption that the gluons are produced
locally in rapidity and keep the same rapidity label as in the
parent distribution function. The first assumption in principle is
easily relaxed by considering $b$-dependent distributions,
although it makes the calculation considerably more cumbersome.
The second assumption is more questionable. One certainly expects
"migration" in rapidity during the interaction, and in particular
the relevant rapidities should depend on the produced transverse
momentum. This is certainly the case in the collinear factorized
perturbative formalism, where the parent rapidities are taken at
$x=k_t/\sqrt s$. However in the present case the production is not
necessarily through the $2\rightarrow 2$ process, and thus the
rapidities are not fixed in the same way. Keeping these caveats
in mind, we now consider the implications of Eq.~(\ref{eq2}).

In a purely perturbative approach, neglecting the effects of
saturation the gluon distribution function at impact parameter $b$
has the shape
\begin{equation}
   \phi_A^{\rm pert}(k_t,b)
   ={\alpha_s(N_c^2-1)\over 2\pi^2}K{\rho_{\rm part}(b)\over 2}
    {1\over( k_t+\Lambda_{QCD})^2}\, ,
   \label{eq4}
\end{equation}
where $\rho_{\rm part}(b)$ is the density of participants, taken as
reference for $A-A$ collisions. The role of $\Lambda_{QCD}$ in the denominator 
is to regulate the gluon distribution in the infrared.
The additional numerical
multiplicative factor $K$ reflects the fact that the gluons at low
$x$ originate not only from the valence quarks, but also from the
sea quarks and energetic gluons.


\section{Models for the gluon distribution}

Saturation effects modify the gluon distribution so that it
is suppressed at low momenta $|k_t|<Q_s$. There are several models
in the literature which provide such saturated distribution
function. In this note we consider two types of models.

\subsection{McLerran--Venugopalan gluon ~\cite{mv}}

The McLerran-Venugopalan model~\cite{mv} achieves saturation by
taking into account the Glauber-Mueller multiple scattering
effects. The intrinsic glue distribution in this model was
calculated in ~\cite{kovchegov,jkmw}
\begin{equation}
  \phi^{\rm MV}_A(k_t)
  ={N^2_c-1\over 4\pi^4 \alpha_s N_c}
  \int {d^2{\bf x} \over {\bf x}^2}
  \left(1-e^{-{\bf x}^2Q_s^2({\bf x}^2)/4}\right)\,
  e^{i\, k_t \cdot {\bf x}}\, .
  \label{eq5}
\end{equation}
 We will
take the saturation momentum to be ${\bf x}$-dependent following
~\cite{kln}
\begin{equation}
    Q_s^2({\bf x}^2,b)
    ={4\pi^2\alpha_s N_c\over N_c^2-1}\,
     xG(x,1/{\bf x}^2) {\rho_{\rm part}(b)\over 2}\, ,
  \label{eq6}
\end{equation}
with $G(x,k_t^2=1/{\bf x}^2)$ being the {\it nucleon} gluon
distribution.  We will take the gluon distribution in the nucleon
to be of the simple perturbative form ~\cite{muller}
\begin{equation}
   xG(x,1/{\bf x}^2)
   = K \ \frac{\alpha_s(N_c^2-1)}{2\pi}
    \ln\left( \frac{1}{ {\bf x}^2\Lambda^2_{\rm QCD}}
              + a \right)\, .
   \label{eq7}
\end{equation}

The small regulator $a = \frac{1}{ {\bf x}_c^2\Lambda^2_{\rm QCD}}$
ensures that the saturation momentum stays positive for ${\bf x}^2
\gg {\bf x}_c^2$. For the numerical evaluations we choose $x_c =
3~ {\rm GeV}^{-1}$, such that for momenta $k_t \ge O(1 ~{\rm GeV})$ 
the sensitivity on the infrared cut off $a$ is negligible. We take
$\Lambda_{\rm QCD} = 0.2~{\rm GeV}$, and $\alpha_s = 0.5$. 
 The saturation scale $ Q_s^2(b)$ is obtained from the
solution of the implicit equation Eq.~(\ref{eq6}), when evaluated
at ${\bf x}^2 = 1/  Q_s^2(b)$. At $b = 0$ we fix $Q_s^2 = 2~ {\rm
GeV}^2$ throughout this paper; this corresponds to $K=1.8$ in
Eq.~(\ref{eq7}).

\begin{figure}
\hspace*{-0.4cm}\includegraphics[width=84mm]{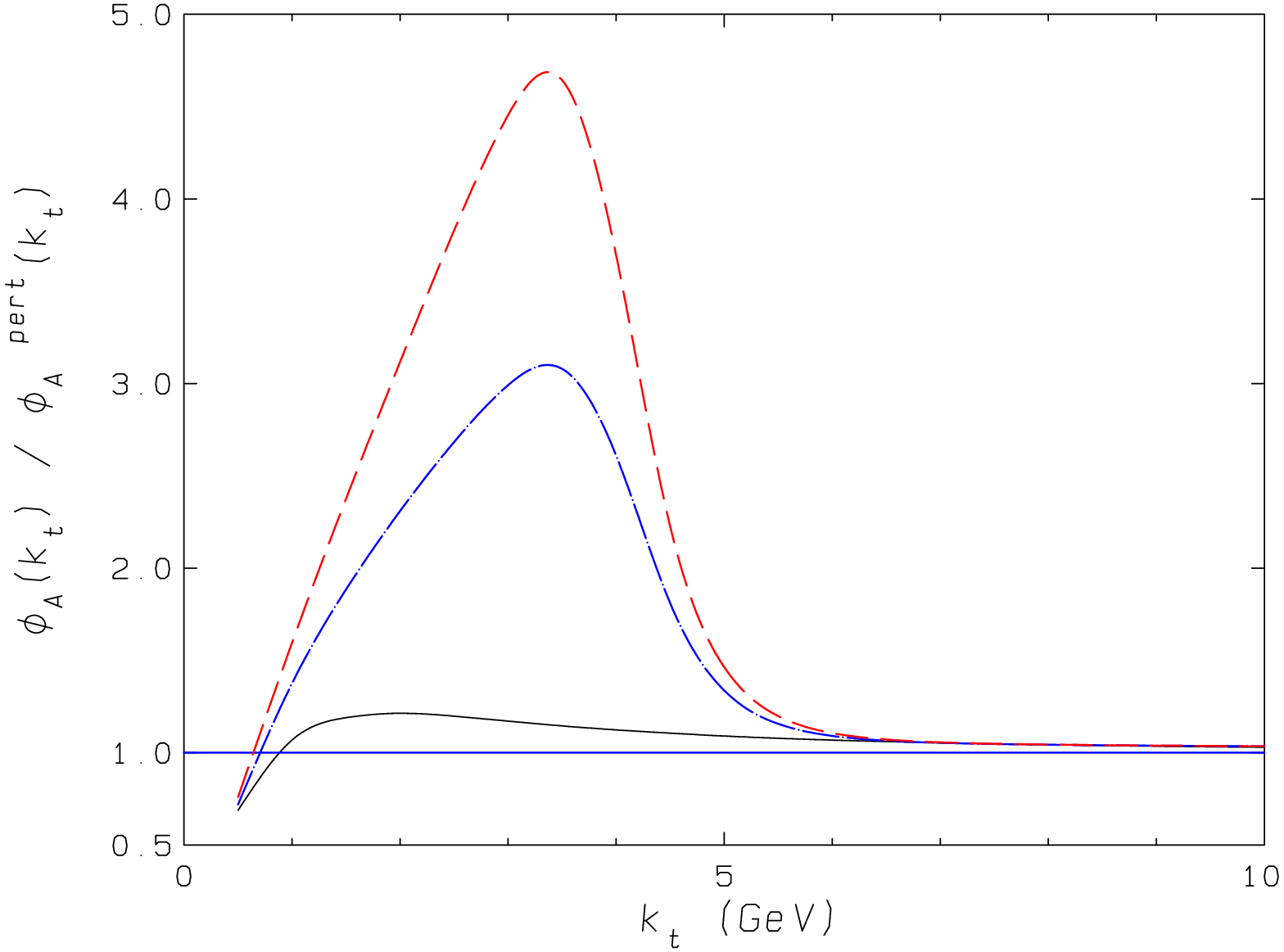}
\hspace*{-0.2cm}\includegraphics[width=84mm]{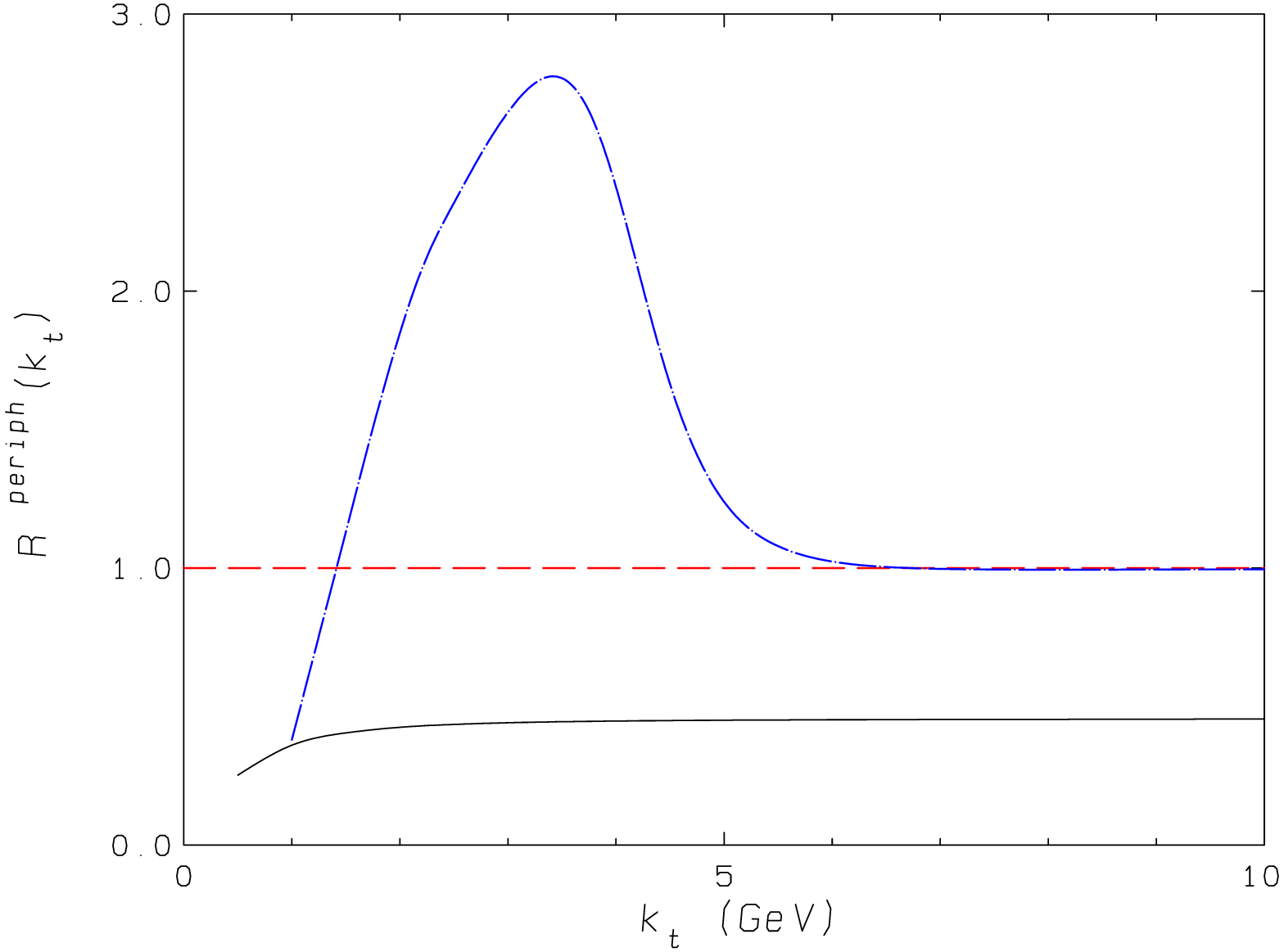}
\caption{a). The unintegrated gluon distribution function, normalized to 
the corresponding perturbative one, as function of $k_{t}$ for fixed 
$\rho_{\rm part} = 3.1~ {\rm fm^{-2}}$ and for $Q_s^2 \simeq 2~ {\rm GeV}^2$; 
solid curve: gluon distribution in the McLerran-Venugopalan model ~\cite{mv}, 
dot-dashed curve: model (\ref{eq10}) - (\ref{eq12}) for the evolved
gluon distribution. The dashed curve is for the anomalous dimension
$\gamma = 0.5$. b). The ratio $R^{\rm periph}$ in Eq.~(\ref{rperiph}) 
for the gluon distributions Eq.~(\ref{eq10}) (dot-dashed curve) and 
Eq.~(\ref{slow}) (solid curve).
}
\label{fig:Fig1}
\end{figure}

The MV distribution appears in the light cone gauge calculation as
the average of the gluon number density operator in the state with
random distribution of color charges distributed with the nuclear
density ~\cite{jkmw}. On the other hand as shown in
~\cite{kovchegov}, in the covariant gauge calculation of DIS-like
processes it accounts precisely for the rescattering of the
produced gluon inside the nucleus. The effect of these multiple
rescatterings is to broaden the gluon transverse momentum spectrum
by the amount $\Delta k^2_t\sim Q_s^2$. Accordingly, the low
momentum part of $\phi^{\rm MV}_A$ is suppressed relative to the
perturbative gluon, the region around $k_t\sim Q_s$ is enhanced,
while at large momenta $k_t\gg Q_s$ there is no appreciable change
relative to the perturbative  gluon. Fig.\ref{fig:Fig1}a shows
the ratio of the MV gluon to the perturbative one as a function of
momentum for fixed coupling constant $\alpha_s=0.5$. The multiple
rescatterings do not change the total number of gluons but only
redistribute the gluon momentum. Thus the gluon distribution
$G(Q^2)$ calculated with $\phi_A^{\rm MV}$ is the same as the
perturbative one for $Q^2\gg Q_s^2$.

In a DIS-like process with a probe directly coupled  to gluons,
the intrinsic momentum distribution would be directly proportional
to the spectrum of final state gluons. The Cronin effect is
therefore present in the MV gluon {\it ab initio}. This has been
noticed in ~\cite{jmg}. The question we are interested in, is to
what extent this effect shows up in nuclear collisions.


\subsection{Evolved gluons}

The MV gluon distribution does not contain any evolution in $x$.
One way to introduce the $x$ dependence is to adopt the
Golec-Biernat-W\"usthoff procedure ~\cite{gbw}, whereby the
saturation momentum is taken to be energy dependent with the
factor $x^{-\lambda}$, with $\lambda=0.2-0.3$. In this paper we
are not going to explore the energy dependence of the spectrum,
and thus for our purposes the energy independent MV ansatz is
sufficient.

Another type of saturated gluon distribution has been used in
~\cite{klm}. The energy dependence in the RHIC energy range is not
large enough to allow one to explore the perturbatively predicted
$x$ dependence of the distribution function. One can nevertheless
consider $\phi$ (at both 130 and 200 GeV) as being evolved by the
perturbative evolution from lower energies. Such an energy evolution
leaves a distinctive imprint on the $k_t$ dependence of the gluon.
Although the solution of the nonlinear QCD evolution equation
~\cite{nle} has not been analyzed in great detail, its qualitative
features have been discussed in ~\cite{iim}. It has been
argued in ~\cite{iim} that in the wide region of momenta
$Q_s(x)<|k_t|<{Q_s^2(x)\over Q_0}$ the evolved distribution
behaves as
\begin{equation}
   \phi^{\rm NLE}_A(k_t)
   \propto \left[{Q_s^2\over k_t^2}\right]^\gamma
   \label{eq8}
\end{equation}
with the anomalous dimension
\begin{equation}
  \gamma=0.64\, .
  \label{eq9}
\end{equation}
A slightly different analysis of ~\cite{klm} based on doubly
logarithmic approximation suggests $\gamma=0.5$. In either case
due to the large anomalous dimension, $\phi_A^{\rm NLE}$ is very
significantly enhanced over the perturbative $\phi_A^{\rm pert}$ in the
wide range of momenta. At asymptotically large momenta
$\phi_A^{\rm NLE}$ again reduces to the perturbative expression.

As we will see in the following it is important to know how the 
distribution behaves outside the scaling window.
The behavior outside the scaling window is not known analytically.
One expects that at asymptotically large momenta the behavior 
of $\phi_A^{\rm NLE}$ is perturbative, and therefore  
$\gamma_{|k_t|\rightarrow\infty} \rightarrow 1$.
The crossover from the scaling with the 
anomalous dimension to the perturbative one
can in principle be either sharp at the edge of the window, or
can be very gradual and slow. To explore the possible differences between the
fast and slow crossovers we will use two parametrizations 
of the distribution function.

To model the function with the fast crossover we take for illustration
\begin{eqnarray}
  \phi^{\rm NLEF}_A(k_t) = {N^2_c-1\over 4\pi^3 \alpha_s N_c}
   \left( {\frac{{\hat Q}^2_s}{k_t^2 + {\hat Q}_s^2}}
          \right)^{\gamma(k_t)} \, ,
\label{eq10}
\end{eqnarray}
where
\begin{eqnarray}
\label{eq11}
   {\hat Q}_s^2 = {\hat Q}_s^2(b)
    = K~{2\pi\alpha_s^2 N_c} \, {\rho_{\rm part}(b)\over 2}\, ,
\end{eqnarray}
and
\begin{eqnarray}
\label{eq12}
 \gamma (k_t)=
  \frac{1+0.64\, w(k_t)}{1+ w(k_t)}\, ,
  \qquad w(k_t) =  \left( \,
   \frac{10\, Q_s^2(b)}{(k_t  +~\Lambda_{\rm QCD})^2 } \right)^6\, .
\end{eqnarray}
Here, $Q_s^2(b) = 4\, {\hat Q}_s^2(b)$, consistent with Eq.~ (\ref{eq6})
at  $\rho_{\rm part} = 3.1\, \hbox{fm}^{-2}$.

The parametrization of $\phi^{\rm NLE}_A(k_t)$ is chosen to be consistent
with the required behavior at large $k_t$ as well as at
$k_t = O(Q_s)$ ~\cite{iim}. The width of the scaling
window in our parametrization is $\sim 3\, Q_s$.
 For $k_t < Q_s$, the gluon $\phi_A^{\rm NLE}$ is 
supposed to saturate or grow at most logarithmically. 
In our ansatz this saturation is ensured by the presence of the term  
$\hat Q^2_s$ in the denominator of Eq.~(\ref{eq10}).

To model the possibility of the slow crossover we will take simply the function
with fixed anomalous dimension:
\begin{eqnarray}
  \phi^{\rm NLES}_A(k_t) = {N^2_c-1\over 4\pi^3 \alpha_s N_c}
   \left( {\frac{{\hat Q}^2_s}{k_t^2 + {\hat Q}_s^2}}
          \right)^{0.64} \, ,
\label{slow}
\end{eqnarray}
Although this function never approaches the perturbative asymptotics,
for the purposes of numerical evaluation it is indistinguishable from a 
function with slowly varying $\gamma(k)$.

We note that Eq.~(\ref{eq10}) parametrizes the gluon distribution
directly in momentum space. We found that this is the simplest way
to generate an acceptable $\phi$ to be used in the framework of
Eq.~(\ref{eq2}). Alternatively one could try to define $\phi$ via
the frequently used relation involving the dipole scattering cross
section $N({\bf x})$ ~\cite{iim}:
\begin{equation}
  \phi^{\rm NLE}_A(k_t)
  ={N^2_c-1\over 4\pi^4 \alpha_s N_c}
  \int {d^2{\bf x} \over {\bf x}^2}N({\bf x})
  e^{i\, k_t{\bf x}}\, .
\label{eq13}
\end{equation}
where
\begin{equation}
  N({\bf x})=
  \frac{\left({\bf x}^2Q_s^2({\bf x}^2)/4\right)^\gamma}
  {1+ \left({\bf x}^2Q_s^2({\bf x}^2)/4\right)^\gamma} \, ,
\label{eq14}
\end{equation}
with $\gamma (x)$ approaching $\gamma = 0.64$ at small values of
$1/Q_s > x > \Lambda_{\rm QCD}/Q_s^2$ in a way similar to
Eq.~(\ref{eq12}). Although strictly speaking $\phi_A(k_t)$ is
defined by the relation Eq.~(\ref{eq13}) only at large $k_t$,
naively one could expect it to be also reasonable at $k_t\sim
Q_s$. However it turns out not always to be the case. The Fourier
transform in Eq.~(\ref{eq13})
 is not only sensitive to $x \sim 1/k_t$ but to all $x< 1/k_t$. As a result
depending on details of the parametrization of $\gamma(x)$, we
sometimes found "oscillating" unintegrated gluon functions which
for some momenta were even negative. In general, this indicates
that one should be very careful using relation
Eq.~(\ref{eq13}), as even for a reasonable $N(x)$ it can
produce unacceptable $\phi_A^{\rm NLE}(k_t)$.

As seen from Fig.\ref{fig:Fig1}, the qualitative features of the
gluon distribution in the MV model (\ref{eq5}) and the scaling
models (\ref{eq10}) are to some extent similar. When compared to the
leading order perturbative distribution Eq.~(\ref{eq4})
they show suppression at $k_t\ll Q_s$ and enhancement at $k_t >Q_s$. 
In Fig.\ref{fig:Fig1}a, we show the ratio 
\begin{equation}
  R^{\rm pert}=\phi_A(k_t)/\phi^{\rm pert}_A(k_t)\, .
\end{equation}
In the scaling
model the enhancement is much more pronounced. It is clear that
using the same ansatz, but with $\gamma=0.5$ rather than $\gamma=0.64$
would make this enhancement even greater.

In relation to the MV gluon the perturbative distribution Eq.~(\ref{eq4}) 
is the relevant distribution to be used to model the peripheral collisions.
For the evolved functions a more meaningful comparison is with the 
``peripheral'' distribution of the same functional form as $\phi_A^{\rm NLE}$ 
but with a smaller value of $Q_s$.
In Fig.\ref{fig:Fig1}b we plot the ratios 
\begin{equation}
    R_{\rm NLEF(S)}^{\rm periph} 
    = \frac{\phi_A^{\rm NLEF(S)}(b=0,k_t)/\rho_{\rm part}(b=0)}
                      {\phi_A^{\rm NLEF(S)}(b=13\, {\rm fm},k_t)
                        /\rho_{\rm part}(b=13\, {\rm fm})}
      \label{rperiph}
\end{equation}
with $Q_s$ corresponding to $\rho_{\rm part}=3.1\, {\rm fm}^{-2}$ 
and $\rho_{\rm part}=0.35\, {\rm fm}^{-2}$, 
(the values of impact parameters $b=0$ and 
$b=13\, {\rm fm}$) respectively.
We observe that the ratios for $\phi^{\rm NLEF}_A$ and 
$\phi^{\rm NLES}_A$ are very different.
While the ratio $R^{\rm periph}_{\rm NLEF}$ exhibits
enhancement for the momenta in the scaling window,
no such enhancement is seen in $R^{\rm periph}_{\rm NLES}$.
The reason is that with the ansatz $\phi_A^{\rm NLEF}$ as long as $k_t$
is outside the scaling window of the peripheral distribution, 
but inside the window of the central one (that 
is for $3\, Q_s(b=13{\rm fm})<k_t<3Q_s(b=0)$) the ratio 
$R_{\rm NLEF}^{\rm periph}$ 
is practically equal to $R^{\rm pert}$. This is not the case for 
$\phi_A^{\rm NLES}$ because of a very slow approach to asymptotic behavior.
Instead, for all momenta of interest 
\begin{equation}
    R^{\rm periph}_{\rm NLES}=
    \left({Q_s^2(b=0)\over Q_s^2(b=13\, {\rm fm})}\right)^{\gamma-1}
    = {\rho_{\rm part}(b=13\, {\rm fm})\over\rho(b=0)}^{1-\gamma} < 1\ .
\end{equation}
We note that the physics of the enhancement of $R^{\rm pert}$ for 
$\phi_A^{\rm NLE}$ is different from that of
$\phi_A^{\rm MV}$. According to ~\cite{iim} and ~\cite{klm} the
anomalous dimension of the evolved distribution 
is a direct consequence of the linear BFKL (or
doubly logarithmic) evolution and is not significantly affected 
by the nonlinearity
of BK equation.  As opposed to multiple
rescatterings resummed in $\phi_A^{\rm MV}$, the BFKL evolution
greatly increases the total number of gluons relative to the
perturbative distribution. Thus the enhancement of $\phi_A^{\rm NLE}$ is
not due to the gluon number conserving redistribution of the gluon
momentum, but rather due to the BFKL growth of the total number of
gluons. The slow fall off of $\phi_A^{\rm NLE}$ with momentum in the
scaling window and  the associated clear momentum broadening 
is the result of the BFKL diffusion, which fills
the phase space very far from the momentum at which the
distribution is peaked at initial energy.

Although some qualitative features of saturating gluon
distributions appear to be model-independent, quantitative results
can vary significantly, see Fig.\ref{fig:Fig1}a and 
Fig.\ref{fig:Fig1}b. In the rest of
this paper, we discuss the implications of this model-dependence
for the spectrum Eq.~(\ref{eq2}), and we study the behavior of
$dN/d^2p_t$ as a function of transverse momentum and centrality.


\section{Gluon production in A-A collisions}

For the perturbative gluon distribution Eq.~(\ref{eq4}), the main
contribution to the integral in Eq.~(\ref{eq2}) comes from the
region of phase space $k_t-p_t\sim 0$ and $k_t\sim 0,\ \ \
p_t-k_t\sim p_t$. The contribution from the bulk of the phase
space $k_t\sim k_t-p_t\sim p_t$ is logarithmically suppressed due
to the fast decrease of the perturbative distribution function
with $k_t$. One finds
\begin{equation}
  \frac{dN^{\rm pert}(b)}{dy\, d^2p_t}\Bigg\vert_{y=0}
  = 2 S_{AA}(b){N_c^2-1\over 4\pi^3 \alpha_s N_c}
  {{\hat Q}_s^4 (b)\over p^4_t}~
 \left( \ln {p_t^2\over 4 \Lambda^2_{\rm QCD}} + 2 \gamma_E \right) \, ,
 \label{eq15}
\end{equation}
where $\gamma_E$ is the Euler constant. With Eq.~(\ref{eq11})
this spectrum scales with $N_{\rm coll} \propto S_{AA}(b)\,
\rho_{\rm part}^2(b)$ as expected perturbatively.

For the saturated gluon distribution in the MV model the gluon
yield Eq.~(\ref{eq2}) is expressed by
\begin{eqnarray}
  \frac{dN(b)}{dy\, d^2p_t}\Bigg\vert_{y=0}
  = \frac{2}{\pi^3}\, \frac{N_c^2-1}{\alpha_s\, N_c}
  \, \frac{S_{AA}(b)}{p_t^2}\,
  \int_0^\infty \frac{dx}{x^3}\,
 \left(1-e^{-{\bf x}^2Q_s^2({\bf x}^2)/4}\right)^2 \,
 \,  J_0(p_t\, x)\, .
  \label{eq16}
\end{eqnarray}
Since small momenta $p_t$ are suppressed in the MV gluon distribution,
expression (\ref{eq16}) is smaller than the perturbative one. For small 
$|p_t|$, the ${\bf x}$-dependence of
the saturation scale Eq.~(\ref{eq6}) is frozen, and one finds
\begin{eqnarray}
  \frac{dN(b)}{dy\, d^2p_t}\Bigg\vert_{y=0}^{p_t\ll Q_s(b)}
  = \frac{1}{4\pi^3}\, \frac{N_c^2-1}{\alpha_s\, N_c}
  \, \frac{ Q_s^2(b)\, S_{AA}(b)}{p_t^2}\, \ln(4)\, .
  \label{eq17}
\end{eqnarray}
Thus, for small transverse momentum, the spectrum (\ref{eq17})
scales with the number of participants, i.e.
\begin{equation}
  N_{\rm part}(b) = S_{AA}(b)~\rho_{\rm part}(b)\, .
  \label{eq18}
\end{equation}
Numerically, however, we find that the limit (\ref{eq17}) provides a 
fair approximation for the full expression (\ref{eq16}) in a very
small $\rho_{\rm part}(b)$-dependent region below $p_t < 0.1$ GeV only. 

\begin{figure}[t]
\centerline{\vspace*{0.5cm}\epsfig{bbllx=0,bblly=0,bburx=600,bbury=460,
file=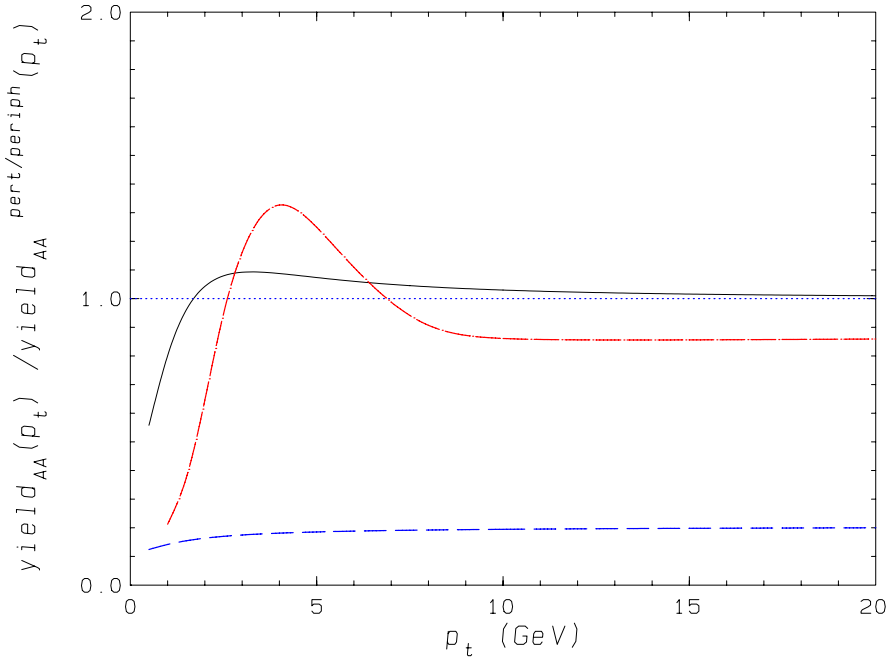,width=120mm}} 
\caption{Cronin effect in the $p_{t}$-dependence of gluon production 
yields for head-on A-A collisions.
Solid curve for the MV-gluon distribution normalized to the perturbative
yield. The dot-dashed curve is for the evolved gluon distribution 
(\ref{eq10}), the dashed one for the evolved gluon distribution 
(\ref{slow}),
both normalized as in Eq.~ (\ref{eq21p}). }
\label{fig:Fig2}
\end{figure}

In Fig.\ref{fig:Fig2} we plot the result of
the numerical evaluation of the formula Eq.~(\ref{eq16})
normalized to the peripheral (perturbative) yield Eq.~(\ref{eq15}), i.e.
\begin{equation}
  \frac{dN(b)}{dy d^2p_{t}}\Bigg /\frac{dN^{\rm pert}(b)}{dy d^2p_{t}} \, ,
  \label{eq19}
\end{equation}
for  $\rho_{\rm part}= 3.1~ ({\rm fm^{-2}})$, and with $~Q_s^2 =  2~
{\rm GeV^2}$. This corresponds to the normalized ratio of central over
peripheral yields 
\begin{equation}
 \frac{1}{N_{\rm coll}(b=0)}
  \frac{dN(b=0)}{dy d^2p_{t}}\Bigg / 
  \frac{1}{N_{\rm coll}(b=13\, \hbox{fm})}
  \frac{dN(b=13\, \hbox{fm})}{dy d^2p_{t}}\, . 
  \label{eq21p}
\end{equation}
To understand the
$p_t$-dependence of (\ref{eq19}) qualitatively, we consider the
integral $\int d^2k_t$ $\phi_A(y,k_t)$ $\phi_A(y,p_t-k_t)$
in (\ref{eq2}). For a saturated gluon distribution and $|p_t|>Q_s$,
one does not gain a logarithmic enhancement
factor from the low momentum region, $|k_t|\ll |p_t|$. However due
to enhancement in the intermediate region one gets a bigger
contribution to the integral from $|k_t|\sim|k_t-p_t|\sim Q_s$.
Since both factors of $\phi$ are enhanced in this region, we
expect that for some momentum range this enhancement will overcome
the suppression in the small momentum range, and therefore will
lead to a net excess of produced gluons relative to the perturbative 
result. This is seen in Fig.\ref{fig:Fig2} as a small but clear Cronin
enhancement of the produced gluon number for momenta just above
the saturation scale. The amount of enhancement depends on the
value of the coupling constant, but qualitatively the phenomenon
persists for any $\alpha_s$.

For the evolved distributions $\phi_A^{\rm NLE}$,
since the $k_t$ dependence is very slow, the contribution
to the integral comes from a very large range of momenta - for
$p_t$ in the scaling region the integral is dominated by
$|k_t|\sim|k_t-p_t|\sim|p_t|$. This leads to a 
significant enhancement of gluon production for all $|p_t|>Q_s$ relative to
the perturbative expression. It is however more interesting to consider 
the ratio of the central to peripheral yields.
Again, taking $\rho_{\rm part}(b=0)=3.1\, {\rm fm}^{-2}$ and 
$\rho_{\rm part}(b=13)=0.35\, {\rm fm}^{-2}$  we display the 
results of the numerical integration of
Eq.~(\ref{eq2}) in Fig.\ref{fig:Fig2}. This plot mirrors 
the plot for a single distribution 
Fig.\ref{fig:Fig1}b. The distribution $\phi^{\rm NLEF}_A$ displays a clear 
Cronin effect similar to the MV gluon, while $\phi_A^{\rm NLES}$ shows 
uniform suppression for the central/peripheral ratio for all momenta.
Although the ratio of the yields for $\phi_A^{\rm NLEF}$ is
smaller than unity at $p_t\sim 10-20 {\rm GeV}$, 
we have checked numerically that at very large $p_t$ it slowly approaches 
unity from below.
We thus conclude that the properties of the ratio are very sensitive to
the way in which the distribution behaves outside the scaling window.

\begin{figure}[t]
\centerline{\vspace*{0.5cm}\epsfig{bbllx=0,bblly=0,bburx=600,bbury=460,
file=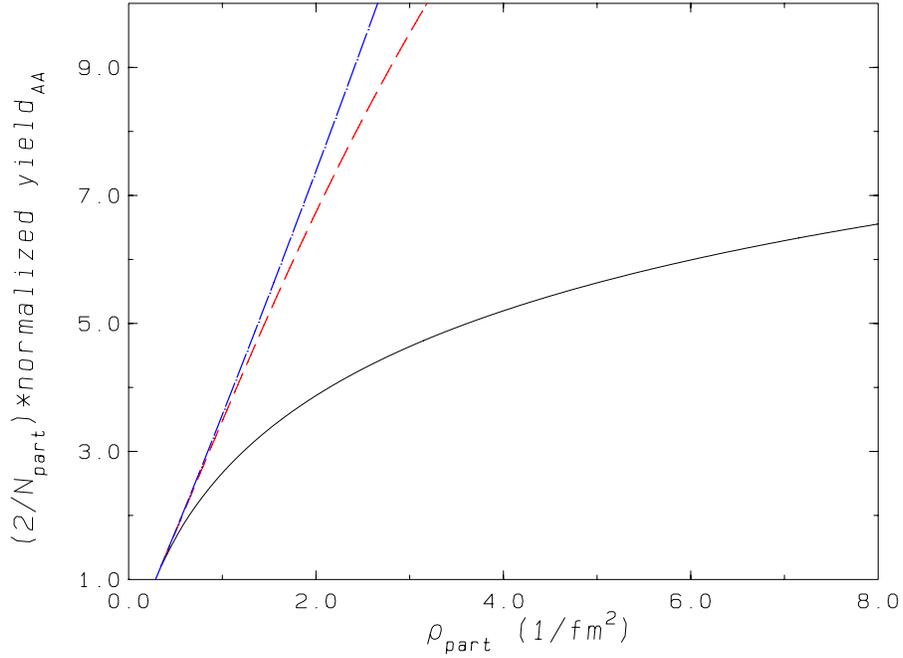,width=120mm}}
\caption{Centrality dependence of gluon production yields (\ref{eq2}) 
in A-A collisions as a function of $\rho_{\rm part}~ ({\rm fm^{-2}})$, 
normalized to the yield in peripheral collisions, Eq.~(\ref{eq20}).
Curves are calculated for the MV gluon distribution (\ref{eq5}) and
different values of $p_t$:
solid curve: $p_{t} = 0.25~{\rm GeV}$, dashed curve:  
$p_{t} = 1.0~{\rm GeV}$, dot-dashed curve:  $p_{t} = 3.0~{\rm GeV}$.
} 
\label{fig:Fig3}
\end{figure}

\begin{figure}[t]
\centerline{\vspace*{0.5cm}\epsfig{bbllx=0,bblly=0,bburx=600,bbury=460,
file=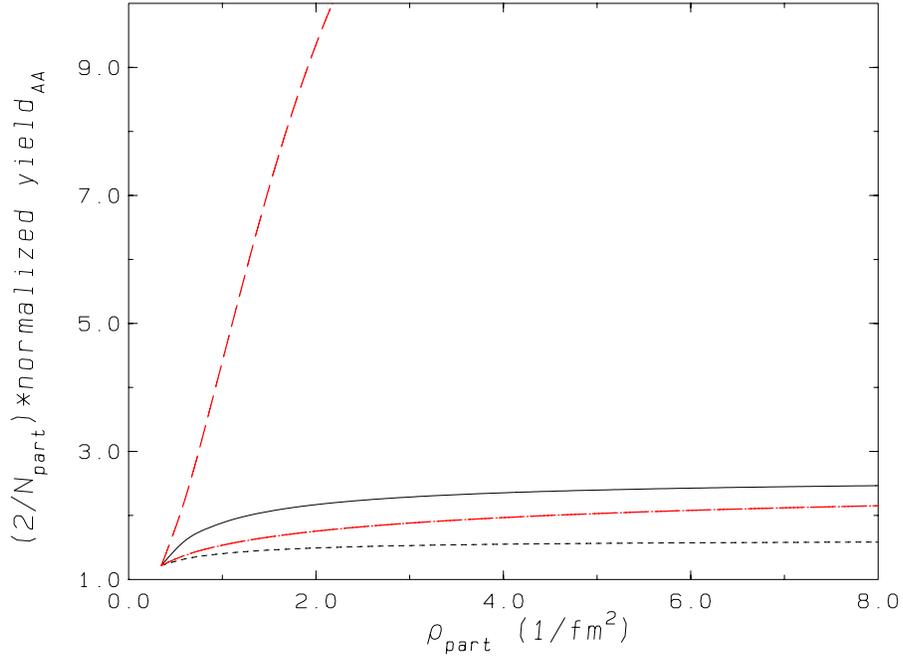,width=120mm}}
\caption{ Centrality dependence as in Fig.~\ref{fig:Fig3}, but
with the gluons of Eq.~(\ref{eq10}) and Eq.~ (\ref{slow}) for
$p_t =$ 1 and 3 GeV.
The solid and dashed lines correspond
to Eq.~(\ref{eq10}) for  $p_t = 1 $ GeV and $p_t = 3$ GeV, respectively. 
The short-dashed ($p_t =$ 1 GeV) and dot-dashed ($p_t = 3$ GeV) lines 
are calculated for the gluon distribution (\ref{slow}).}
 \label{fig:Fig4}
\end{figure}
%
\begin{figure}[t]
\centerline{\vspace*{0.5cm}\epsfig{bbllx=0,bblly=0,bburx=600,bbury=460,
file=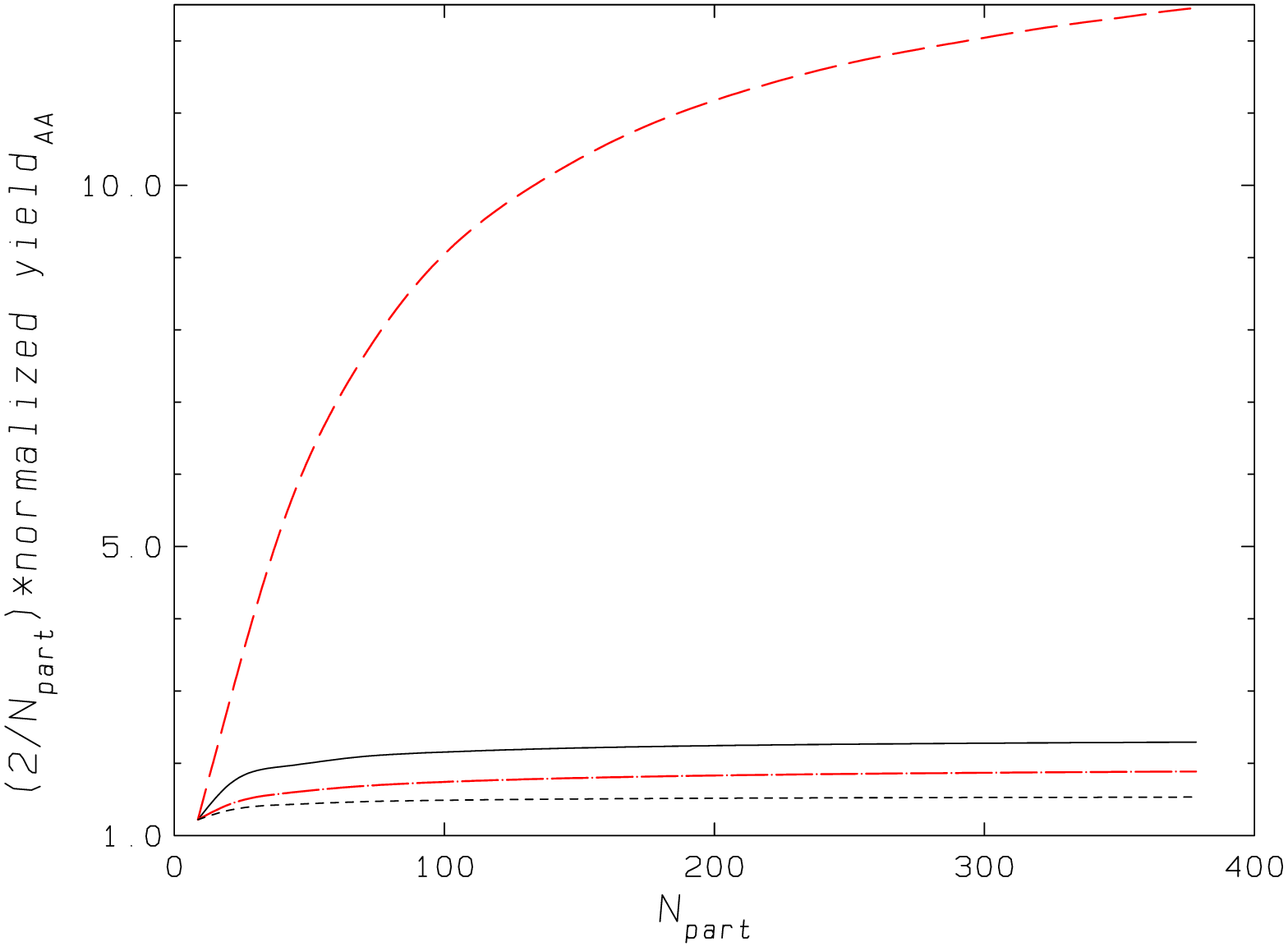,width=120mm}}
\caption{ Normalized yield for evolved gluon distributions 
as a function of $N_{\rm part}$. Legend the same as on Fig.4.
The largest value $N_{\rm part}=380$ corresponds to 
$\rho_{\rm part}=3.1~ {\rm fm}^{-2}$ on
Fig.\ref{fig:Fig4}.} \label{fig:Fig5}
\end{figure}

We discuss now to what extent the gluon spectrum (\ref{eq2}) shows
an approximate $N_{\rm part}$ or $N_{\rm coll}$ scaling in some
kinematic regime. To this end, we plot the gluon yield
normalized to the yield for peripheral collisions at $b = 13.0~
{\rm fm}$, corresponding to $\rho_{\rm part}= 0.35~ {\rm fm^{-2}}$
~\cite{kln},
\begin{equation}
\label{eq20}
   \frac{2}{N_{\rm part}(b)}~\frac{dN(b)}{dy d^2p_{t}}\Bigg /
   \frac{dN(b=13.0)}{dy d^2p_{t}}\, .
\end{equation}
This quantity is plotted for the MV gluon in Fig.\ref{fig:Fig3}
and for the evolved gluon distributions (\ref{eq10}) and (\ref{slow})
in Fig.\ref{fig:Fig4}, for
various fixed values of $p_t$ as a function of $\rho_{\rm part}$. We
also replot in Fig.\ref{fig:Fig5}
this quantity for the evolved gluon as a function of
$N_{\rm part}$ using the relation between $\rho_{\rm part}$ and $N_{\rm part}$
given in ~\cite{kln}. We explore the
dependence on $\rho_{\rm part}$ beyond the experimentally accessible
range to illustrate better the functional dependence. 
For the MV gluon
a steep increase with $\rho_{\rm part}$ indicative of $N_{\rm coll}$
scaling is found for $p_t \ge 1~{\rm GeV}$. For smaller transverse
momentum, e.g. $p_t = 0.25~{\rm GeV}$, the 
$\rho_{\rm part}$-dependence is seen to level
off. 

For the evolved gluon the centrality dependence is again sensitive to the
large momentum behavior. For $\phi_A^{\rm NLEF}$ at $p_t=3\, {\rm GeV}$ the
centrality dependence is similar to the MV gluon. 
One does not recover $N_{\rm part}$ scaling even in the 
enlarged centrality region $\rho_{\rm part}< 8~ {\rm fm^{-2}}$.
The qualitative features of
Figs.\ref{fig:Fig4} and \ref{fig:Fig5} at small and large
$\rho_{\rm part}$ at $p_t=3~{\rm GeV}$ can be understood as follows:
At small $\rho_{\rm part}$, the saturation momentum $Q_s$ is small so
that the scaling window does not exist; the overall yield scales with 
$N_{\rm coll}$, like the perturbative one. This $N_{\rm coll}$
behavior should persist as long as 3 GeV is above the scaling
window, $3\, \hbox{GeV} > Q_s^2(b)/Q_0$ 
(for our parametrization $Q_0=0.5\, {\rm GeV}$). This is indeed seen
for small $\rho_{\rm part}$ in Figs.\ref{fig:Fig4} and \ref{fig:Fig5}.
In the other extreme, when $Q_s^2$ becomes very large and 3 GeV lies
well inside the scaling window, the main contribution to
the yield comes from the integration over the momentum in the
scaling window and one obtains
\begin{equation}
  \frac{dN(b)}{dy d^2p_{t}}\propto S(b)Q_s^{2\gamma}\, . 
  \label{eq21}
\end{equation}
This dependence is much flatter. For $\gamma=0.5$ as in
~\cite{klm}, it scales with $N_{\rm part}$ rather than
$N_{\rm coll}$. For the value $\gamma = 0.64$ used here, 
the dependence is only slightly steeper. This is the argument
given in ~\cite{klm}. However, as we have shown above, at these 
large values of $Q_s$ the absolute magnitude of
$\frac{dN(b)}{dy d^2p_{t}}$ is greater than that of
$\frac{dN^{\rm pert}(b)}{dy d^2p_{t}}$. Hence there must
be an intermediate region of $\rho_{\rm part}$ where $\frac{dN(b)}{dy
d^2p_{t}}$ grows {\it faster} than $N_{\rm coll}$ so that
$\frac{dN(b)}{dy d^2p_{t}}$ can overtake $\frac{dN^{\rm pert}(b)}{dy
d^2p_{t}}$ and then stay flat for some region of large
$\rho_{\rm part}$ \footnote{Of course
$\frac{dN(b)}{dy d^2p_{t}}$ does not stay above 
$\frac{dN^{\rm pert}(b)}{dy d^2p_{t}}$ for arbitrarily large 
$\rho_{\rm part}$. When the value of $Q_s$ reaches $p_t$,  
$\frac{dN^{\rm pert}(b)}{dy d^2p_{t}}$ overtakes
$\frac{dN(b)}{dy d^2p_{t}}$ consistently with our discussion of
the previous section.}. Such behavior is indeed seen in
Figs.\ref{fig:Fig4} and \ref{fig:Fig5}. However, the values of
$\rho_{\rm part}$ for which the curve flattens out, are
larger than the experimentally relevant ones, see
Fig.\ref{fig:Fig5}. The reason why the scaling of
Eq.~(\ref{eq21}) is not reached faster is that the argument
leading to Eq.~(\ref{eq21}) neglects the finite width of the
scaling window. Since the function $\phi_A^{\rm NLEF}$ decreases quite
slowly in the scaling window, the yield gets significant contributions 
from momenta above $Q_s^2/Q_0$. These momenta bring in additional
$Q_s$ and therefore  $\rho_{\rm part}$ dependence, and slow down the
onset of scaling Eq.~(\ref{eq21}). 
At smaller momenta ($p_t=1{\rm GeV}$) 
the normalized yield is flat in $\rho_{\rm part}$ for 
$\rho_{\rm part}>2\, {\rm fm}^{-2}$, since these gluons are produced below 
the saturation momentum.

For $\phi_A^{\rm NLES}$ on the other hand, the centrality dependence 
is very flat for all momenta we explored, consistent with~\cite{klm}.

\section{Gluon production in p-A or d-A collisions}

Finally, we discuss the gluon production in the situation where the
distribution of one of the nuclei is perturbative. This situation
pertains to proton-nucleus and
deuteron-nucleus collisions. Following the previous discussion, we
use the $k_t$-factorized formula (\ref{eq2}) for the gluon yield,
replacing $\phi_A(y,k_t)$ $\phi_A(y,p_t-k_t)$ in
(\ref{eq2}) by the product of a perturbative gluon distribution
for the proton and a saturated MV or evolved gluon distribution
for the nucleus. We also compare the results of this calculation
to the gluon production cross section derived in the
quasi--classical approximation ~\cite{km,kw,kovchegov}. This quasi
classical expression can be written as
\begin{eqnarray}
 \frac{d \sigma^{pA}}{d^2p_t \ dy \ d^2b}
  &=& \frac{1}{\pi} \, \int  \, d^2 {\bf x}\,
      d^2{\bf y}  \frac{1}{(2 \pi)^2}  \frac{\alpha_s C_F}{\pi}
      \frac{{\bf x} \cdot {\bf y}}{{\bf x}^2 {\bf y}^2} \,
      e^{i p_t\cdot ({\bf x}-{\bf y})} \,
  \nonumber \\
   && \times
   \left[ \left(e^{ - ({\bf x}-{\bf y})^2 \, Q_s^2 / 4 } - 1 \right) +
          \left( 1 -  e^{- {\bf x}^2 Q_s^2 /4 } + 1 - 
               e^{- {\bf y}^2 Q_s^2 /4 } \right) \right]\, ,
    \label{eq22}
\end{eqnarray}
where $b$ is the impact parameter, which we choose to be  $b=0$,
$p_t$ is the gluon's transverse momentum, and $y$ its
rapidity. Although this expression also has a factorized form in
 momentum space, this form is distinct from
Eq.~(\ref{eq2}) ~\cite{kt}. For the numerical evaluation, we
regulate the ${\bf y}$-integration of the first bracket of
(\ref{eq22}) by an infrared cut-off $1/\mu$ where we choose $\mu
= \Lambda_{\rm QCD}$. This cut-off can be introduced using $\ln
\frac{1}{{\bf z}^2 \mu^2} \, = \, \frac{1}{\pi} \, \int d^2{\bf y}
\, \frac{{\bf y} \cdot ({\bf z} + {\bf y})}{{\bf y}^2
 ({\bf z} + {\bf y})^2}$. As in Eq.~ (\ref{eq6}), we use for the
saturation scale
\begin{eqnarray}
   Q_s^2 ({\bf x}) \ =  \frac{4 \pi^2 \alpha_s N_c}{N^2_c - 1}\, 
   (2 R \rho) \, xG (x, 1/{\bf x}^2)\, ,
   \label{eq23}
\end{eqnarray}
where $\rho$ denotes the nuclear density and $xG$ is the gluon
distribution of Eq.~ (\ref{eq7}). For the numerical analysis, we
use $2 R \rho = \rho_{\rm part}(b=0)/2 = 1.5 ~{\rm fm}^{-2}$.

\begin{figure}[t]
\centerline{\vspace*{0.5cm}\epsfig{bbllx=0,bblly=0,bburx=600,bbury=460,
file=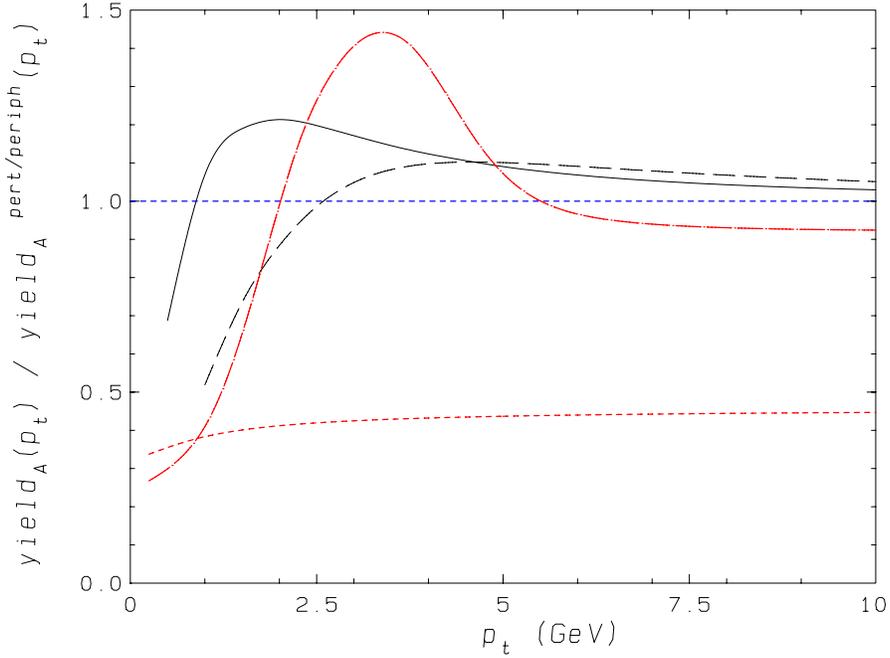,width=120mm}} \caption{The gluon yield produced in
p-A collisions, normalized to the perturbative yield for the MV
gluon and to the peripheral yield for the evolved gluon distributions. The
different curves are for the quasi-classical expression (\ref{eq22})
normalized to (\ref{eq29}) (dashed curve), the $k_t$-factorized spectrum 
with MV gluon (solid curve), and with evolved gluon
distributions (\ref{eq10}) (dot-dashed curve) and Eq.~ (\ref{slow})
(short-dashed curve), respectively.} \label{fig:Fig6}
\end{figure}

In order to compare with the perturbative behavior, we calculate
for central rapidity $y = 0$ the asymptotic limit of
(\ref{eq22}),
\begin{eqnarray}
   \frac{d \sigma^{pA}}{d^2p_t \ dy \ d^2 b }
   \Bigg\vert_{{p_t} \gg Q_s}
  = \frac{4 \alpha_s^3 \ C_F}{\pi} \ K  \ (2 R \rho) \,
  \frac{1}{p_t^4} \, (\ln [\frac{p_t^2}{4 \Lambda_{\rm QCD}^2}]
  +2 \gamma_E - 1)\, .
  \label{eq29}
\end{eqnarray}
In the limit of small momenta, we quote ~\cite{km}
\begin{eqnarray}
  \frac{d \sigma^{pA}}{d^2p_t \ dy \ d^2 b }
  \Bigg\vert_{{p_t} \ll Q_s} =
  \frac{\alpha_s \ C_F}{\pi^2} \ \frac{1}{p_t^2}\, ,
  \label{eq30}
\end{eqnarray}
which is obtained for "frozen", i.e. constant $Q_s$.
We use Eq.~(\ref{eq29}) as the 
baseline for comparison for the MV gluon. For 
the evolved gluon we calculate the yield using the standard
factorized formula. We take the proton distribution in the same
functional shape as that of the nucleus, i.e. Eq.~(\ref{eq10}) and 
Eq.~(\ref{slow}) with
$\hat Q_s=\Lambda_{QCD}$. We  compare the central yield ($\rho_{\rm part}=
3.1\, {\rm fm^{-2}}$) to the peripheral yield 
($\rho_{\rm part}=0.35\, {\rm fm^{-2}}$). 
Results are plotted in Fig.\ref{fig:Fig6}.

For moderate momenta, i.e. above about twice the
saturation scale $Q_s \simeq 1.4~ {\rm GeV}$, the full rate for 
the MV gluon shows a Cronin-type
enhancement with respect to the perturbative one. For small
momenta $p_t \le Q_s$, there is significant suppression, which can
be immediately deduced when comparing (\ref{eq30}) with
(\ref{eq29}). A qualitatively similar behavior is found when
calculating the gluon spectrum from the factorized ansatz
(\ref{eq2}). As in the case of the nucleus-nucleus collisions, we
find also that in p-A collisions the Cronin ratio in the 
evolved case depends strongly on the properties of the evolved 
distribution above the scaling window. 


\section{Conclusion}

In summary, our study illustrates that  quantitative results
depend largely on the precise model-dependent implementation of
saturation effects. A generic qualitative feature for the 
multiple scattering situation is that
perturbative saturation leads to Cronin-type
transverse momentum broadening of the produced gluon spectrum in
both nucleus-nucleus and proton(deuteron)-nucleus collisions.
In the evolved case 
there is an overall enhancement of the production yield relative
to the perturbative baseline. The spectrum also exhibits strong
transverse momentum broadening due to relatively large anomalous dimension.
On the other hand,
the absence or presence of the Cronin effect strongly
depends on the behavior of the distribution function outside the
scaling window. At present the shape of the evolved distribution in this
momentum range is not known analytically. The recent numerical study 
~\cite{anticronin} strongly indicates that crossover from the scaling
regime to the perturbative one is very slow and gradual, and that the Cronin
effect which is present in the MV gluon is wiped out by the quantum evolution
at high energies. Thus, $\phi_A^{\rm NLES}$ in Eq.~(\ref{slow}) seems
to provide a more realistic parametrization of the evolved gluon
distribution than $\phi_A^{\rm NLEF}$ in Eq.~(\ref{eq10}).

Although the centrality dependence of the produced gluon spectrum
shows $N_{\rm part}$ scaling in some limiting case, it can differ
significantly from a simple $N_{\rm part}$ or $N_{\rm coll}$
scaling in the experimentally accessible regime depending on the 
shape of the gluon distribution .

A detailed comparison of perturbative saturation models
to data not only requires the knowledge of the distribution and
the improvement of the calculation of the gluon production yield
beyond the factorized expression (\ref{eq2}) used in this paper.
It also requires the inclusion of fragmentation functions
for gluons into pions, and a discussion of their possible
medium-dependence~\cite{sw}. 

On the qualitative level however we observe that the gluon distributions 
which lead to the Cronin effect in d-Au collisions also lead to the Cronin
enhancement in the Au-Au collisions. And vice versa, if no Cronin effect 
appears in Au-Au, none is seen in d-Au collisions.
Given the recent experimental observation of the Cronin enhancement
in d-Au collisions at RHIC ~\cite{dA} this supports the view that
significant final state (``quenching'') effects are needed in
order to account for the Au-Au data ~\cite{phenstar}.

{\bf Note added.}
When preparing the revised version of this paper we were made aware of 
~\cite{kkt} and ~\cite{jnv} which also study the effects of saturation on the 
Cronin enhancement. These references agree with our results regarding
the MV gluon. Regarding the evolved gluon, the detailed numerical study
is reported in ~\cite{anticronin}.

\subsubsection*{Acknowledgments}

The authors thank the Institute for Nuclear Theory at the
University of Washington for its hospitality and the Department of
Energy for partial support during the completion of this work.
We thank A.~H.~Mueller for clarifying remarks and important 
suggestions when revising this work. We also thank E.~Levin for
asking the right questions.
Useful discussions with N.~Armesto, Y.~Kovchegov, P.~Jacobs,
C.~A.~Salgado and D.~Schiff are gratefully acknowledged. 
R.~B. is supported, in part, by DFG, contract FOR 329/2-1. A.~K. is
supported by PPARC Advanced Fellowship.


\end{document}